%% file: pap5AAv5.tex
\documentclass[twocolumn]{aa}
\usepackage{graphicx}
\usepackage{epsfig}
\usepackage{natbib}
\def\pmb#1{\setbox0=\hbox{#1}%
 \kern-.025em\copy0\kern-\wd0
 \kern.05em\copy0\kern-\wd0l
 \kern-.025em\raise.0433em\box0}
\def\eg{{\it e.g.\ }}

\def\ie{{\it i.e.\ }}

\newcommand{\no}{\noindent}                                                       
\newcommand{\EQ}{\begin{equation}}
\newcommand{\eq}{\end{equation}}
\def\zmax{$z_{max}\;$}
\newcommand{\EQA}{\begin{eqnarray}}
\newcommand{\eqa}{\end{eqnarray}}

\input{macros}

\begin{document}

\title{Sensitivity and figures of merit for dark energy supernova surveys
}

%\subtitle{Evolution of the non-linear galaxy bias up to z=1.5}

\author{
J.-M. Virey \inst{1},
A. Ealet \inst{2}
}

%\offprints{J.-M. Virey, e-mail: virey@cpt.univ-mrs.fr}

\institute{
Centre de Physique Th\'eorique\thanks{``Centre de Physique Th\'eorique'' is UMR 6207 
- ``Unit\'e Mixte
de Recherche'' of CNRS and of the Universities ``de Provence'',
``de la M\'editerran\'ee'' and ``du Sud Toulon-Var''- Laboratory
affiliated to FRUMAM (FR 2291)},
Case 907,
F-13288 Marseille Cedex 9, France\\
and 
Universit\'e de Provence, Marseille, France
\and
Centre de Physique des Particules de Marseille,
Case 907
F-13288 Marseille Cedex 9, France
}

%\authorrunning{Virey J.-M. and Ealet A.}
%\titlerunning{Sensitivity and figures of merit for dark energy supernova surveys}

\date{July, 2006}

\abstract{

Tracking the origin of the accelerating expansion of the Universe   
remains one of the most challenging research activities today. The  
final answer will depend on the precision and on the consistency of  
data from future surveys. The sensitivity of these surveys is related to 
the control of the cosmological parameters errors. We focus on supernova surveys 
in the light of the figure of merit defined  
by the Dark Energy Task Force. 
We estimate the impact of the level of systematic errors on the
optimisation of SN surveys and emphasize their importance deriving
any sensitivity estimation. We  discuss the lack of
information of the DETF figure of merit to discriminate among
dark energy models and compare the different representations
that can help to distinguish $\Lambda$CDM from other
theoretical models.
We conclude that all representations should be controlled
through combined analysis and consistency checks to avoid biases.


}

% Preprint CPT-P55-2006
\maketitle

\section{Introduction}

The discovery of the acceleration of the Universe is one of the most intriguing questions in astrophysics today and
has been the driver of many theoretical developments trying to find 
an explanation for this acceleration. These models introduced in general a new component called
dark energy (``the DE models'') whose nature is unknown \citep[see \eg][]{pee,pad,cop}. 
Their comparison to observational data is 
complex and the experimental interpretation would benefit from a 'model independent approach'. 
The strategy is not unique today and is linked to the definition of the cosmological parameters,
in particular those that describe the properties of the dark energy component. 

Many studies concentrate on the equation of state ($w=$ pressure/density ratio) of this new component.
For a cosmological constant, $w$ is equal to $-1$  but it can be different and/or can vary with time
in other DE models. A common way to introduce the time dependence of DE models 
is to use a redshift dependent parameterization such as \citep{lin03,pol01} :
\EQ
w(z)=w_0+w_az/(1+z)=w_0+(1-a)w_a
\eq 
$a$ being the scale factor. 
The function $w(z)$ is a good observable with adequate properties
\citep{lin04a,lin05}.

On the theoretical side, one needs to estimate whether this parameterization is 
sufficient to describe DE models whatever the source of the acceleration is. 
This has been investigated by  \citet{lin06a,barg,lin04b} 
who show that this parameterization  can represent
a large class of models, even those that have no real dynamical component. Nevertheless, 
there are still some potential biases when estimating the effective $w_0$ and $w_a$
parameters from the actual theoretical phase space \citep{lin06a,brid}.

On the experimental side, many new probes have shown their ability to constrain $w(z)$.
To estimate the sensitivity of an experimental survey,  named "a data model", 
one needs to define figures of merit (FoM), related to the statistical 
error of the $w(z)$ parameters. This can be used to compare and test different experimental strategies. \\

The aim of this article is to compare the information provided by the various FoM
calculated for Supernova (SN) surveys. In particular, we distinguish FoM needed to compare data models from FoM needed to distinguish between and separate DE models. In section 2, we define the generic SN surveys used as data models in this article.
In section 3, we recall the definition of the FoM used by the Dark Energy Task Force \citep{DETF}
and use it to compare the sensitivities of the data models. We optimize the total number of SN,
$N$, and the redshift depth, $z_{max}$, of the survey in light of the systematic errors.
In section 4, we compare the impact of different FoM to distinguish  the source
of the cosmic acceleration among the DE models.

\section {Generic supernova data models}

The sensitivity of future surveys depends on several experimental parameters. 
One concern is to well estimate their
uncertainties. For this purpose, we concentrate here on Supernova surveys. 
We define generic SN data models that are representative of future data :\\

\no a)  $N=2000$ and $z_{max}=1$: such a survey is close to what can be reached 
from the ground in the near future. Using the DETF terminology, this will be our 
definition of a "stage 2" data model.\\

\no b)  N=15000 and $z_{max}=1$: this is achievable from the ground 
with wide coverage. 
We call it a "stage 3" data model
or a "wide" survey.\\

\no c)  N=2000 and $z_{max}=1.7$: this needs an infrared coverage which implies a space mission. This will be possible at a later
stage and we will define it as our "stage 4"
or a "deep" survey.\\

\no d)  N=15000 and $z_{max}=1.7$: this is postponed to the future and  we will define 
it as a  "stage 5" or a "wide and deep" survey.\\

 Comparison of the potentialities of theses data models should describe the expected 
improvement on the constraints of the DE equation of state with the characteristics of the survey.
We focus in particular on the relative importance of increasing the total number of SN ($N$) collected
against the redshift depth ($z_{max}$). The feasibiliy of a deep or a wide survey  
is strongly correlated to the experimental strategy and is a driver of future surveys. 
To address this, we will keep the other parameters identical for all the data models. 
We use the following hypothesis for the cosmology:
\begin{itemize}
\item a $\Lambda$CDM model, 
\item a flat universe,
\item a strong $\Omega_m$ prior, as expected from Planck : $\Omega_m =0.27 \pm 0.01$. 
The central value has been
chosen to be in agreement with the WMAP-3 year data analysis.
\end{itemize}
We add the following assumptions:
\begin{itemize}
\item we add a sample of nearby supernovae as expected from the SN Factory 
survey \citep{SNFactory} corresponding to 150 SN at z=0.03 and 150 at 0.08. We call
it the ``nearby sample'' in the following.
\item The intrinsic magnitude dispersion is assumed to be 0.15. 
The corresponding "statistical error" of a redshift bin is 
$\delta m_{stat}=0.15/\sqrt{N_{bin}}$. We have assumed 
redshift bins of width $\Delta z=0.1$.
%\item the magnitude dispersion is defined as 0.15 per redshift bin of width $\Delta z=0.1$, 
%corresponding to a "statistical error" of 
%$\delta m_{stat}=0.15/\sqrt{N_{bin}}$ per redshift bin.
\end{itemize}

The level of systematic errors will appear as a fundamental ingredient in the data model comparison (it is true for any analysis as it has been emphasized by \citet{DETF} and \citet{kim}). 
We define the systematic error by an extra term 
$\delta m_{syst}$  in the magnitude.
The total error on the magnitude $m(z)$ is then $\delta m^2=\delta m_{stat}^2+\delta m_{syst}^2$.
Unless otherwise specified, we use an uncorrelated error in redshift bins.
We have also estimated the effects of a redshift dependent error (\ie correlated in redshift bins) with different amplitudes. 
Adding a redshift dependence does not change our 
conclusions on errors. It is the amplitude of the systematic errors, 
whatever the form, that has a strong impact on the future precision \citep{kim}.

We study the two cases with and without this systematic term. Since many papers still provide analysis with statistical errors only, we start with  this assumption ($\delta m_{syst}=0$) in our study.
We then choose a default value of
$\delta m_{syst}=0.02$ for the systematic case. This choice is an optimistic estimation of the experimental systematic errors.
The motivation is that using 2000 SN, this systematic error value is already roughly of the same size as 
the statistical error and is a limiting factor of the total error.

Higher systematic error values, more realistic, provide similar conclusions when compared to the statistical case. Only the parameter error values are different.

These two ``academic'' scenarios illustrate the impact of a small systematic effect 
when statistical errors are at a percent level.\\

To perform the simulations we adopt a standard Fisher matrix approach which allows a rapid estimate
of the parameter errors following the procedure described in \citet{vir}. 
We use the freely available tool ``Kosmoshow''\footnote{``Kosmoshow'' is available 
at http://marwww.in2p3.fr/renoir/Kosmoshow.html}.\\

The redshift distribution used in this study is based on the SNAP prescription from~\citet{kim}.
The "stage 4" data model has exactly this distribution. For other data models with
different $N$ and/or $z_{max}$  we have scaled the distribution, truncating it at
$z_{max}=1$ when relevant and multiplying the remaining number of SN by the adequate factor
\footnote{For example, for "stage 2", we take the initial
distribution ($N=$2000 SN up to $z_{max}=1.7$) up to $z_{max}=1$, this selects 1171 SN
then we multiply the number of SN in each redshift bin by 2000/1171 to obtain the
desired $N=2000$ total number of SN. Then, we get the stage 3(5) distribution by multiplying
by 7.5 the number of SN of the stage 2(4) distribution.}.

\section{ Supernova data model sensitivity }

We study the potential of the previous four generic data models 
in term of coverage and statistics, with and without systematic errors.
We examine the interpretation 
of the pivot redshift and of the FoM defined by the DETF. The impact
of a systematic error on the sensitivity of these surveys is emphasized. 
Finally, we give some insights
to the optimization of $z_{max}$ and $N$.

\subsection {The DETF figure of merit}

Recently, the DETF\citep{DETF} has proposed a FoM derived from the definition of the pivot
point. The pivot parameterization is defined as :
\EQ
w(z)=w_p+(a_p-a)w_a=w_p+\frac{w_a}{1+z_p}-\frac{w_a}{1+z}
\eq 
and is equivalent to the parameterization given in eq.1. 
The pivot redshift $z_p$ is defined by \citep{hu}:

\EQ
z_p=-C_{w_0w_a}\sigma (w_0)/(\sigma (w_a)+C_{w_0w_a}\sigma (w_0))
\eq
where $C_{w_0w_a}$ 
is the correlation between $w_0$ and $w_a$ and
$\sigma (w_i)$ is the error on the parameter $w_i$.

It has been shown \citep{alb} that 
the ($w_0$,$w_a$) and ($w_p$,$w_a$) contours are mathematically equivalent.
In fact, $w_p$ is directly related to $w_0$ and $w_a$ through a linear transformation : 
$w_p= w(z_p)=w_0+w_az_p/(1+z_p)$. Consequently, any volume in phase space is conserved. 
This change of definition
is  convenient to determine the mathematical redshift $z_p$  
where the function $w(z)$ has the smallest  statistical  error since parameters are decorrelated 
(this corresponds to the so-called ``sweet-spot'', see \eg \citet{hut}).
Note that $z_p$ has been shown to be analysis dependent and has then no real physical meaning 
\citep{lin06b,alb}. 
Note also that the error on $w_p$ is equivalent to the 
one obtained on $w$ when we fix a constant $w$ (as often done 
in previous work in the literature) since $w_p$ and $w_a$ are decorrelated.\\

The DETF FoM is defined as  $[\sigma(w_p)\times\sigma(w_a)]^{-1}$ \citep{DETF}
and is proportional to the inverse of the area of the error ellipse enclosing
the 95\%CL in the $w_0$-$w_a$ plane.
In the following, we 
use this DETF FoM or a ratio of it, 
where the normalization  can change from case to case.
We study and show how this FoM can compare data models. We then examine how this FoM 
can help to discriminate DE models (see also \citet{lin06b}).

\subsection {Evolution of $z_p$}

Figure 1 gives the evolution of $z_p$
for the four data models under consideration, with and without systematic errors. 
$z_p$
ranges from 0.14 to 0.27 and is sensitive to the systematic errors and the data model. 
We have also studied the variation of 
 $z_p$ when changing the SN nearby sample, 
the $\Omega_m$ prior or  the fiducial cosmology.
We find, for instance, that the statistics of the SN nearby sample or 
the $\Omega_m$ prior have a stronger impact on the $z_p$ value than the variation
of $z_{max}$.
We conclude that $z_p$ varies in the range 0 to 0.5 and the variations are 
neither physical nor intuitive.
This study has confirmed  that this parameter 
is not representative of any physical characteristic
of a survey (or of the DE dynamics) and should not be used for any comparison.\\

%\begin{figure}[h]
%\begin{tabular}[h]{c c}
%\centerline{\subfigureA{\psfig{file={fig1a.eps},height=8truecm}}
%\subfigureA{\psfig{file={fig1b.eps},height=8truecm}}}
%%\subfigureA{\psfig{file={f6_b.eps},width=8truecm,height=10truecm}}}
%\end{tabular} 
%%\vspace{-1cm}
%\caption{\footnotesize On the left, evolution of $z_p$ for the data models with (black/solid)
%and without (red/dashed) systematic errors. On the right,
%DETF FoM, normalized to the stage 2
%data model, with (black/solid) and without (red/dashed) systematic errors.}
%    \label{fig:fig1} 
%\end{figure}

\begin{figure}
\centering 
\includegraphics[width=8.7cm, angle=0]{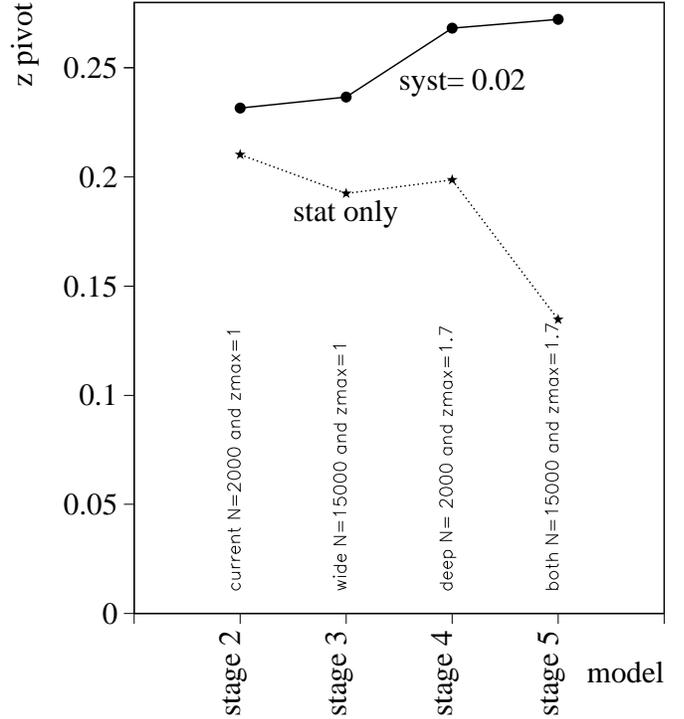}
\caption{Evolution of $z_p$ for the data models with (bullets)
and without (stars) systematic errors.
%Evolution of $z_p$ for the data models with (black-plain curve)
%and without (red-dashed curve) systematic errors.
}
\label{fig1} 
\end{figure}

$w_p$ then  can be ambiguous as corresponding to different redshift values for each data model.
The contours in a plane where $w_p$ is one of the two variables (\eg the $(w_p,w_a)$ or
$(\Omega_m,w_p)$ planes) will be more difficult to 
interpret and should be taken with
caution when comparing surveys with different characteristics. 
However,  there is no ambiguity with the DETF FoM, as it corresponds to the area of 
the error ellipses which is identical  in the planes $(w_p,w_a)$ and $(w_0,w_a)$ \citep{DETF}.
Similarly, $w_p$ can be used to exclude a cosmological constant when compared to $-1$
(but other observables are maybe more efficient, see section 4).

%%%%%%%%%%%%%%%%%%%%%%%%%%%%%%%%%%%%%
%\MTfig{zp}{zpstatsyst.eps}{7.}{Evolution of $z_p$ for the data models with (black/solid)
%and without (red/dashed) systematic errors.}
%%%%%%%%%%%%%%%%%%%%%%%%%%%%%%%%%%%%%%%%%%%%

\subsection {The DETF figure of merit for the SN data models}

The  DETF FoM for the four data models are shown in 
figure 2. 
Stars correspond to calculations with statistical errors only and 
bullets when a systematic error of 0.02 is included.

\begin{figure}
\centering 
\includegraphics[width=8.7cm, angle=0]{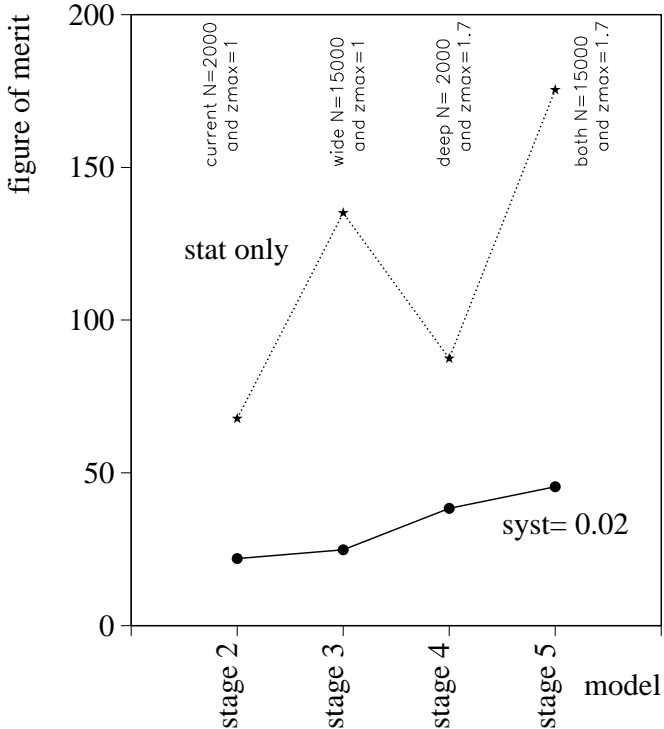}
\caption{DETF figure of merit
%, normalized to the stage 2 data model, 
with (bullets) and without (stars) systematic errors.
%DETF figure of merit, normalized to the stage 2
%data model, with (black-plain curve) and without (red-dashed curve) systematic errors.
}
\label{fig2} 
\end{figure}

The strong impact of the systematic error term appears very clearly in figure 2.
First, we observe a strong reduction of the FoM value when
systematic errors are included.
Secondly, we see different variations of the two cases, when increasing the size of
the survey ($N$) or the survey depth ($z_{max}$). \\

With statistical errors only, a large statistical sample 
yields a better FoM than a deep survey as can be seen in the stages 3 and 4.
The relative improvement of stage 3  to stage 2 is of the order of 2.0,
whereas it is only 1.3 for stage 4 relative to stage 2. In other words, a stage 3 (wide survey)
provides a  sensitivity 55\% better than  a stage 4 
(deep survey).
This indicates that it is better to increase $N$ rather
than \zmax after stage 2.

Concerning the stage 5 data model, there is  also a strong improvement of the
FoM. It increases by 30\% (100{\%}) compared to the stage 3 (4)
data model.
\\

With systematic errors, the conclusion is reversed and stage 4 has a far better potential
than stage 3.
We see in figure 2 an improvement of the order of 56\% for stage 4 relative to stage 3.

The relative improvement of stage 3 to stage 2 is now only of the order of 1.15 compared to 1.75
for stage 4 to stage 2, showing it is now preferable
to increase \zmax rather than $N$ after stage 2. 

Concerning stage 5, the improvement relative to stage 4 is only of 18\% 
which is moderate compared to the technical complexity of such surveys.
\\

For other values of systematic errors, these results are confirmed :
\begin{itemize}
\item With systematic errors higher than $\delta m_{syst}=0.02$,
the difference
between stages 3 and 4 steepens (\ie the FoM ratio increases)
which reinforces the need for a deep survey.
\item Stages 3 and 4 are equivalent, if both have some 
systematic errors of the order of 0.006. Below this value a wide survey
(stage 3) is better, and above it a deep survey (stage 4) is preferable.
\item If we keep $\delta m_{syst}=0.02$ for stage 4, 
then the systematic errors of stage 3 should be controlled
at a better level than 0.012 to be more efficient than stage 4.
\item Using a redshift dependence for the error gives
similar conclusions, only the various quoted values will slightly change.
\item The relative merit of stage 5 compared to stage 4 
or stage 3
is very dependent on the level of systematic
errors assumed in the analysis.
\end{itemize}

The latest SN data from \citet{Riess06} can be interpreted 
as an  improvement  of the FoM of the order of 5.
Then a stage 2 or 3 survey with a level of systematic $\delta m_{syst}=0.05$
will not improve the current result and may be considered as useless.  
Only an improvement 
of the systematic level can help. The limitation  of stage 4 and 5
is a systematic error of 8{\%}.\\

Consequently, the control of the level of systematic errors is the key parameter to discriminate 
between future wide and deep SN surveys. This is understandable 
since the statistical error soon
will be dominated by these systematic errors.
Our conclusions are valid to quantify the impact of systematic 
errors but give no estimate on the methods needed to derive such a level of control. 
A large sample should help in understanding systematic errors and its 
size will depend essentially on the SN properties. 
In the next section, we perform a more detailed analysis of the 
optimisation between $N$, \zmax  and the systematic errors.

\subsection{Optimization of the survey depth }

\citet{linhut} have emphasized
that ``the required survey depth depends on the rigor of our scientific investigation''.
They show that it is mandatory to have $z>1.5$ to reduce cosmological and DE models 
degeneracies when systematic error are included to avoid wrong precision and biased results.\\

In this section, we optimize  the depth $z_{max}$ using the figure of
merit on data models that have a fixed number of SN but a different $z_{max}$. 
We  move $z_{max}$ by steps of size $0.1$ equivalent to
the redshift bin size of our SN distribution.

For each adjacent model, we compute the ratio of the FoM defined for the model at 
\zmax to the one at an adjacent redshift bin  of $z_{max}-0.1$ :
\EQ
R=\frac{FoM^{z_{max}}}{FoM^{z_{max}-0.1}}
=\frac{(\sigma (w_a)\sigma (w_p))^{z_{max}-0.1}}{(\sigma (w_a)\sigma (w_p))^{z_{max}}}.
\eq

In figure 3, we plot this ratio for the two statistics $N=2000$ and $N=15000$, 
with and without systematic errors.
If adding a new redshift bin does not improve the errors,  the ratio is close to one.
The gain is defined by the difference to 1.

\begin{figure}
\centering 
\includegraphics[width=8.7cm, angle=0]{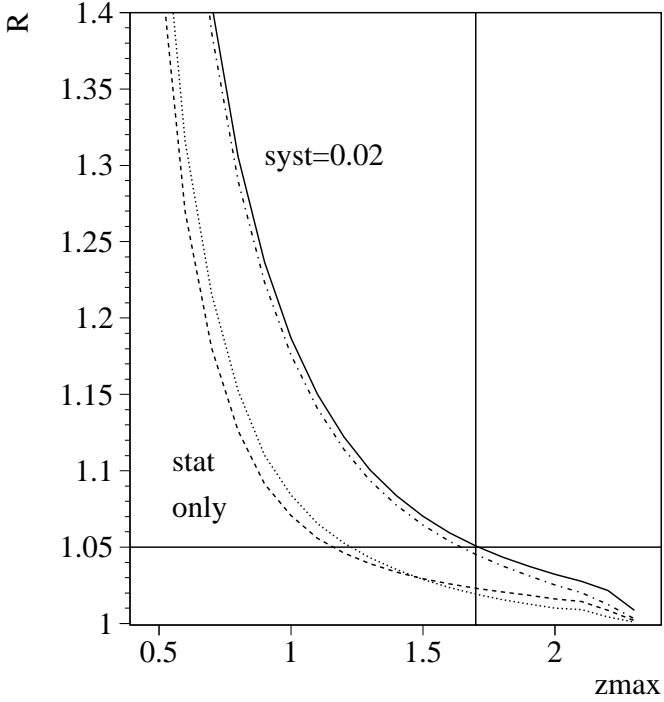}
\caption{Evolution with $z_{max}$ of the
DETF FoM ratio assuming either $N=2000$ or $N=15000$ SN and for
the two cases with and without systematic errors (plain (dash-dotted) 
curve : systematic case with 15000 (2000) SN,
dashed (dotted) curve : statistical case with 15000 (2000) SN). 
The FoM ratio is defined for adjacent $z$-bin (see text).}
\label{fig3} 
\end{figure}

The four curves of Fig.3 show the same behaviour and a large improvement 
is seen when increasing $z_{max}$,
up to a plateau where the gain start to be small.
If we take a 5\%  gain (the horizontal line in Figure 3) as the minimal improvement we can accept 
for an increase of the survey, we can estimate an optimum for $z_{max}$: \\
- with systematic errors of 0.02 and $N=15000$ (plain curve) 
%(black/plain curve) 
 $z_{max}=1.7$ (the vertical line corresponds to $z_{max}=1.7$); \\
- with systematic errors of 0.02 and $N=2000$ 
(dash-dotted curve) 
%(blue/dash-dotted curve) 
the change is small and $z_{max}\approx 1.65$;\\
- with statistical errors only  and $N=15000$ (dashed curve) \zmax is strongly reduced at $1.15$; \\
- with statistical errors only  and $N=2000$ (dotted curve) one gets $z_{max}\approx 1.25$. \\
 
Surprisingly, the statistical case has a relatively small dependence on $N$.
This comes from cancellations in the $z_{max}$ evolution of
the $w_a$ and $w_p$ constraints, which exhibits strong variations with $N$ but
in opposite directions.\\

This can be better understood by plotting directly the FoM (not the ratio).  
Figure 4 shows that the FoM increases with $z_{max}$ and  also with $N$.
With systematic errors, the FoM is not strongly dependent on  $N$ whereas  
with statistical errors only, the variations due to $N$ are stronger than the ones due to $z_{max}$. 
For example, one has the same FoM (90) for N=2000 with $z_{max}=1.7$ 
and for N=15000 with $z_{max}=0.7$. \\

\begin{figure}
\centering 
\includegraphics[width=8.7cm, angle=0]{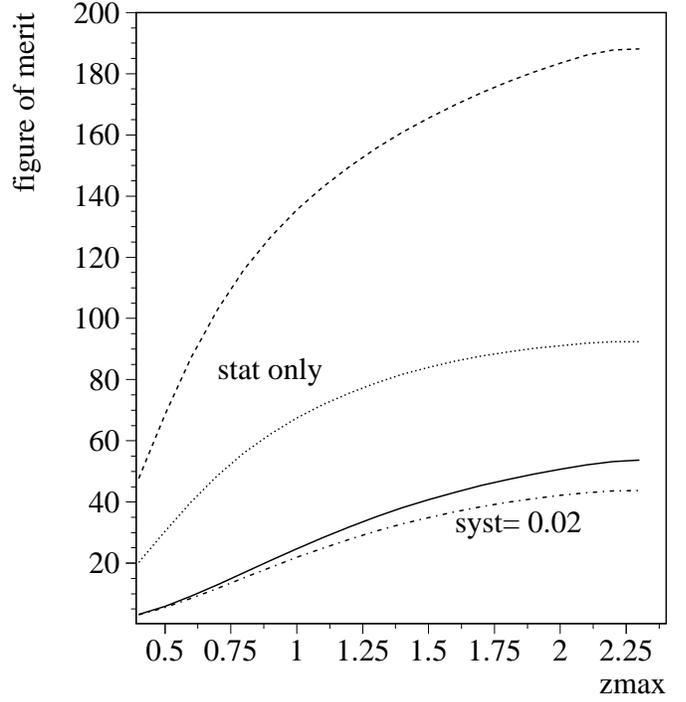}
\caption{The DETF figure of merit (unnormalized) as a
function of $z_{max}$ with the same labels as fig.3.}
\label{fig4} 
\end{figure}

 We deduce from this study :
\begin{itemize}
\item When the systematic errors are neglected, $N$ is the fundamental parameter and $z_{max}$
around 1 is sufficient to derive strong constraints on the dark energy equation of state.
\item If systematic errors are of the order of $\delta m_{syst}=0.02$, a SN sample at $z>1$ is mandatory
to increase the constraints
that can be reached from the ground (\eg stages 2 and 3) whatever the statistical size.
\item Beyond $z>1.7$, the improvement is marginal (less than 5\% by redshift
bin of size 0.1). This result is weakly dependent on $N$ but dependent on $\delta m_{syst}$.
\item A survey with a higher level of systematic errors requires a higher $z_{max}$ and 
has a reduced dependency on $N$.
\item If we introduce systematic errors with a redshift dependency 
(a correlation between bins), 
the conclusion is identical to the constant case and the optimal $z_{max}$ 
depend on the larger systematic error value.   
\end{itemize}

Thus, we see that for any realistic surveys, it will be more important to 
have a coverage beyond z=1 to be able to control the precision than to increase 
the statistics. This is an important driver for any future SN survey.

\section {Comparing Dark Energy models}

We want to address not only 
the statistical sensitivity of the SN surveys but also their capability to separate Dark Energy models.
More precisely, we would like to know if a particular DE model is in 
agreement with the standard $\Lambda CDM$ model. 
Then we need to test the compatibility of the two models.
 The  DETF FoM is only one number
and does not allow us to answer this question. 
For example, Linder has shown \citep{lin06b} that the key discriminant 
for thawing (freezing) models (see \citet{cal}
for model definitions) is the long (short) axis of the ($w_0$,$w_a$) ellipse . 

Consequently, contour plots should provide more information than the DETF
FoM and/or the pivot point to interpret
data.
We have looked in more detail at the different information to estimate the most useful FoM when we compare 
DE models.\\
The information is contained in the following  FoM: $\sigma (w_0)$, $\sigma (w_a)$,
$\sigma (w_p)$, $\sigma (w_0)\times\sigma (w_a)$, $\sigma (w_p)\times\sigma (w_a)$, 
the two-dimensional ($w_0$,$w_a$) contours  and the redshift function $w(z)$
with its error shape $\sigma (w(z))$. 
We do not consider contours with $w_p$ since they are mathematically equivalent to the contours
with $w_0$ instead \citep{alb}, and multiple $w_p$ contours are difficult
to interpret (see section 3.2).\\

The study of the variations of the different errors, as we have done in the previous section for the
DETF FoM, is particularly interesting in comparing data models. The individual variations
of $w_0$, $w_a$ and $w_p$ do not provide any supplementary information to that given by
the DETF FoM. The behaviours of $\sigma (w_0)$, $\sigma (w_a)$ and the DETF FoM are very similar.
Only $\sigma (w_p)$ behaves very differently as it has a very weak dependence on data models, in particular
it has almost no dependence on \zmax (see fig.3 of \citet{linhut}).

To compare DE models, in addition to the errors  we also need the central
values of the cosmological parameters. 
We focus now on two different FoM : contour plots in the ($w_0$,$w_a$) plane 
and some representation of $w(z)$ with error shape variations with the redshift.\\

%Figure 3 shows  95\% CL contours in the ($w_0$,$w_a$) plane with and without systematic errors
%for the stage 3 and 4 data models. We recover the same conclusion than systematic errors 
%inverse the comparision of the two surveys. The problem of these contour plots is that
%we cannot put too many different contours for the figure to be understandable. T he 4 surveys with and without
%systematic will give a not reachable figure ! Then comparing numbers is in this case 
%far more easier. However, as mentionned earlier, one drawback of the FoM\ in terms of numbers
%is that it is difficult to compare DE models which is much easier on contours.

Figure 5 gives the 95\% CL contours in the ($w_0$,$w_a$) plane for the four data
models with systematic errors. We see two sets of contours, the larger ones with the data models
with $z_{max}<1$ and the smaller ones with $z_{max}>1$, a result already obtained from the DETF FoM.
(Adding in this figure the curves corresponding to the pure statistical cases for the four data models
allows us to recover the results of section 3.3, however the resulting figure is not easily readable.)\\

%%%%%%%%%%%%%%%%%%%%%%%%%%%%%%%%%%%%%
%\MTfig{fig3}{fig3.eps}{7.}{95\% CL contours in the ($w_0$,$w_a$) plane 
%for the four data models with systematic errors. Stage 2 corresponds to the black contour,
%Stage 3 to the green one, Stage 4 to the red one and Stage 5 is in blue.
%%Stage number is given on each contour.
%}
%%%%%%%%%%%%%%%%%%%%%%%%%%%%%%%%%%%%%%%%%%%%
\begin{figure}
\centering 
\includegraphics[width=8.7cm, angle=0]{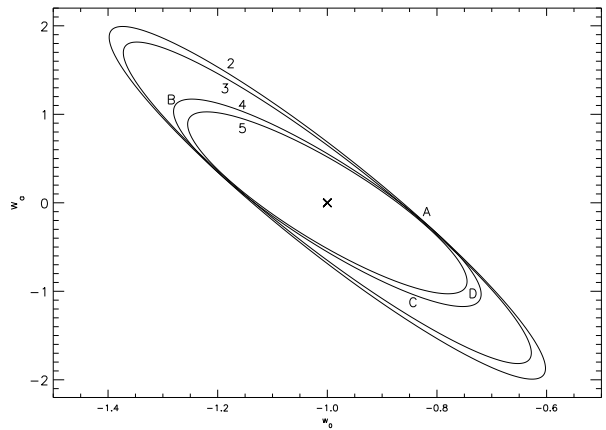}
\caption{95\% CL contours in the ($w_0$,$w_a$) plane 
for the four data models with systematic errors. The labels
of the DE models are given on the plot
(with a small shift of model D for clarity).
Stage number for the data models are given on each contour.
%Stage 2 corresponds to the black contour,
%stage 3 to the green one, stage 4 to the red one and stage 5 is in blue.
}
\label{fig5} 
\end{figure}

The advantage of the ($w_0$,$w_a$) contour representation is the possibility to
``directly'' define some classes of DE models in this plane. Several recent works have been devoted
to this subject (Barger et al., 2006; Linder, 2006a; see also Caldwell \& Linder, 2005;
Scherrer, 2006; Chiba, 2006, for DE model trajectories/locations in the ($w$,$w'=dw/dlna$)
plane).
%\citep{barg,lin06a} \citep{cal,lin06a,barg,sche,chib}.
Consequently, in fig.5 we can represent the different 
classes of models and study their compatibility
with $\Lambda CDM$ in each data model. 
In this way, Linder has shown \citep{lin06b} that to increase
the constraints on ``thawing'' (``freezing'') models we need to reduce the long (short) axis of the ellipse.
However, to optimize such constraints we need to know which parameters control the long and short axis
(and their directions) of the contour. To understand what is important in a survey
to improve the discrimination among DE models, we introduce four phenomenological
models defined by their ($w_0$,$w_a$) pair of values (see Table 1) and which are at the boundary of
the 95\% CL of the stage 4 data model. We study the variation of the constraints for these DE models
for the four data models.\\

%{\small 
\begin{table}[htb]
\begin{center}
\begin{tabular}{|l|c|c|c|c|c|c|c|c|c|c|c|c|c|c|c|c|}
\hline
DE model & $w_0$ & $w_a$ & parameters & $z$-region \\
\hline
A & $-0.82$ & $-0.2$ & $w_p$ & $z_p$ \\
B & $-1.27$ & 1.15 & $w_0$,$w_a$ & low-$z$ \& high-$z$\\
C & $-0.85$ & $-1$ & $w_a$ & high-$z$\\
D & $-0.75$ & $-0.95$ & $w_0$ & low-$z$\\
\hline
\end{tabular}
\caption{\sl Definition and properties of the DE models taken for illustration.
The ``parameters'' (``$z$-region'') column gives the cosmological parameters 
(the redshift region) which are the most efficient
to distinguish the DE model from a cosmological constant.}  
\label{tab:tab1}
\end{center}
\end{table}
%}
%Model A, with parameters $w_0^A=-0.82$, $w_a^A=-0.2$, is closed
%to the border of the thawing region and, from fig.3, we see it will be difficult to exclude 
%this model even with stage 5!
%Conversely,  better is the survey , better is the exclusion of the models B, C and D defined with 
%$w_0^B=-1.27$, $w_a^B=1.15$, $w_0^C=-0.85$, $w_a^C=-1$, $w_0^D=-0.75$, $w_a^D=-0.95$. 
Model A is close
to the border of the thawing region and, from fig.5, we see it will be difficult to exclude 
this model even at stage 5.
On the other hand, the exclusion of models B, C and D is improved with better surveys.

It appears that
the  sensitivity to the data models is higher along the larger axis. 
Consequently, the optimization of future SN surveys are able to reduce the degeneracy
among $w_0$ and $w_a$ which is represented by a reduction of the long axis, whereas it has almost no
impact on the short axis. This conclusion should be tempered. Indeed, if all the ellipses
for the data models meet in two points (model A being one of them) this 
is mainly due to two assumptions :
we have taken the same SN distributions (including the same nearby sample) 
and the same systematic errors
($\delta m_{syst}=0.02$). An important effect of these assumptions concerns 
the orientation of the ellipses
but realistic survey characteristics allow only small rotations of the contours. 
The strongest effect, as mentioned in the
previous section, comes from the assumed level of systematic errors 
which is the fundamental limitation of the size of
the constraints. Then to go beyond,
a better understanding of the systematic errors is mandatory.\\
% The use of
%combined analysis of other cosmological probes (as CMB, WL or ...) will provide also a way to control 
%systematical effects }.\\

It is possible to represent the result in a different way which can be easier  for theoreticians.
The error on $w(z)$, within the Fisher matrix approximation,  is given by :
\EQA
\sigma^2 (w(z))=\sigma^2 (w_0)&+&\sigma^2 (w_a)\frac{z^2}{(1+z)^2}\nonumber\\
               &+&2C_{w_0w_a}\sigma (w_0)\sigma (w_a)\frac{z}{(1+z)} \; .
\eqa

Figure 6 represents  the equation  of state $w(z)$ for the different DE models with the error shapes
obtained for the various data models.

%%%%%%%%%%%%%%%%%%%%%%%%%%%%%%%%%%%%%
%\MTfig{fig4}{fig4.eps}{7.}{Reconstructed w(z) and the associated errors
%for the four data models with systematic errors. The errors correspond by increasing order to
%stage 5, 4, 3 and 2, respectively. The red, green, blue and pink dotted curves give the
%equation of state for models A, B, C and D, respectively.
%%The dotted curves give the
%%equation of state for models A-D which are labelled on the plot.
%}
%%%%%%%%%%%%%%%%%%%%%%%%%%%%%%%%%%%%%%%%%%%%
\begin{figure}
\centering 
\includegraphics[width=8.7cm, angle=0]{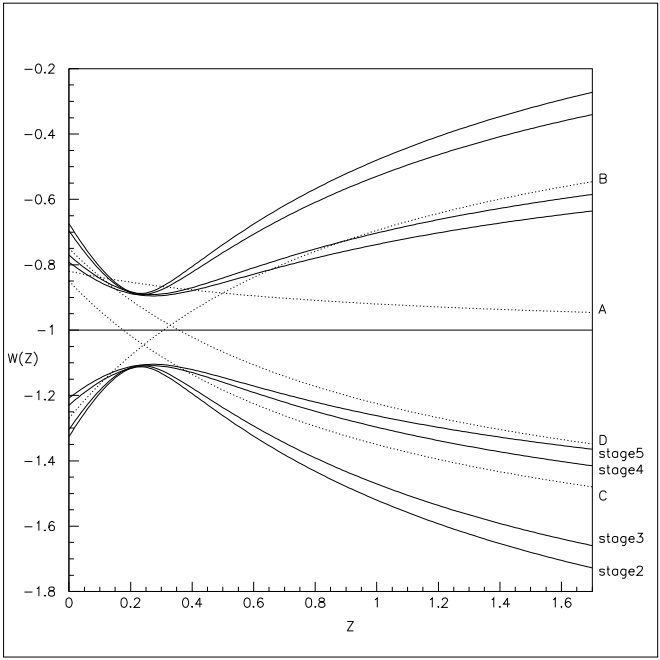}
\caption{Reconstructed $w(z)$ and the associated errors
for the four data models with systematic errors. The errors correspond in increasing order to
stage 5, 4, 3 and 2. 
The dotted curves give the
equation of state for the models A-D and are labelled on the plot.
%The red, green, blue and pink dotted curves give the
%equation of state for models A, B, C and D, respectively.
}
\label{fig6} 
\end{figure}

In this figure, we represent the expected error at 2 $\sigma$ 
of each survey for a $\Lambda$CDM fiducial model
compared with the true $w(z)$ of each DE models (A-D). This representation 
is complete
and explicitly gives the $z$ dependence of the constraints.
The four DE models are described with the same parameterization and show behaviours outside the
2 $\sigma$ limit of the error shape of the $\Lambda$CDM model for stages 4 and 5. They are excluded 
 for different reasons:
\begin{itemize}
\item model A is excluded by the best constraints at the pivot redshift, \ie the best
observable to exclude A is $w_p$.
\item model B is excluded by the low $z$ and high $z$ constraints.
\item model C is excluded by the high $z$ (\ie $w_a$) behaviour.
\item model D is excluded by the low $z$ (\ie $w_0$) behaviour.
\end{itemize}

The  $w(z)\, vs\, z$ representation has the advantage to visualise
 the existence of the sweet-spot at the pivot point and its impact on the result. Anyway, it shows
also that it is not possible to use $w_p$ only.  
For example, model A is excluded by the $w_p$ constraints at $z=z_p$ and not
thanks to the SN discovered at this redshift. This is an example of the difficulty 
of using this information
as physical.
Then, even if this representation is convenient 
its should be taken with some caution.
The error shapes given by eq.4 are strongly parameterization dependent. Consequently, some bias
may be present if the chosen parameterization ($w(z)=w_0+w_a z/(1+z)$ in this study) is far from reality.
In addition, there are strong correlations among the cosmological parameters  and among the redshift bins, 
and the error shapes of $w(z)$ may have some artefacts if not used 
in a realistic redshift range (\ie the range probed by data).\\

Nevertheless, beside the above difficulties this representation may be useful for consistency checks.
This representation is also sensitive to the 
systematic errors and whatever the parameterization is,  one can express the constraints and 
make some data model comparisons. We emphasize that 
it is also possible to provide some results in this plane
that are independent of any choice of
parameterization to describe the DE dynamics, like the so-called ``kinematical'' approach 
\citep[see \eg][]{daly03,daly04}. 
This kind of analysis has 
also some problems of interpretation: errors are in general
difficult to estimate and more noisy, and 
 it does not avoid the correlation in redshift bins of the results.
However, these various approaches are complementary and may be confronted in this plane.\\
 
Thus, excluding particular DE models from a cosmological constant,
require the use of the ($w_0$,$w_a$) plane and/or of the $w(z)\, vs\, z$ representation.
%If our goal is to constrain a particular class of DE models, the ($w_0$,$w_a$) contours are
%the most convenient since the DE model classes are ``well'' defined in this plane.
This is far better than simply comparing $w_p$ with $-1$.
In order to obtain more subtle details, like the connection to a particular class of DE models
or the $z$ dependence of the constraints, both representations are useful.
For instance,
the expression of
the redshift dependence has some advantages in breaking the degeneracy line present 
in the ($w_0$,$w_a$) plane. Models along this
line may be discriminated from a cosmological constant by the measurement of the low and high redshift
behaviour of the equation of state, as encoded in the $w_0$ and $w_a$ parameters.
But DE models that are orthogonal to the degeneracy line may be excluded by the 
constraints at the
pivot redshift, whose expected precision depends weakly on the SN survey configuration
but more on the control of systematic errors.\\
The expected interpretation is then very dependent on all the details of the design of the SN surveys, and in
particular very dependent on the level of systematic errors. This strategy is also dependent on
the chosen parameterization, whose effect should also be carefully estimated.

\section{Conclusions}

We have studied, using the DETF figure of merit, the optimisation
and interpretation of future supernova surveys compared to the forthcoming
ground precision.

We find that the DETF figure of merit is a good approach for testing  the optimisation of a survey.

We test this approach by looking at the sensitivity of the surveys in  
term of the number of SN and on the depth of the survey with particular attention to
the effects of systematic errors. 
%We  emphasize that conclusions are driven by the control of systematic  errors 
%which fix the amount of information inside data. Ignoring them  can provide strong biases on 
%the fitting value and increasing  statistic will only reduce the statistical error on a wrong 
%cosmology.  This conclusion is not depending on the distribution of the  systematic errors.
%
%Then, for an optimistic case of 2\% of uncorrelated systematic errors, we show that  
%there will be no extra information for the cosmology with a statistic larger than 2000 SN and that 
%the  gain will mainly come from an increase of the depth  of the survey up
%to a redshift of 1.7. This optimisation is very 
%strong and only change for a  statistical error only scenario.\\
The DETF figure of merit is very powerful to show the 
difference in sensitivities of surveys with large 
statistics compared to deep surveys with smaller 
statistics. We show, for example, that adding 1 or 2\% of 
systematic errors changes drastically the optimisation and push 
to increase the depth rather than the number of objects. This 
conclusion is very strong when not only statistical errors
are considered, and is not dependant on the kind of systematic
errors we can consider (\eg  correlated in redshift or not).

More precisely, for 2\% of uncorrelated systematic errors, we 
show that there will be no extra information for the cosmology 
with more than 2000 SN and that the  gain will 
mainly come from an increase of the depth  of the survey up to 
a redshift of 1.7.\\

The drawback of the DETF method is the lack of information to estimate  
the discriminating power among DE models, as the central values of  the parameters are not used.

Contour plots in the $(w_0,w_a)$ plane give a better understanding and a good discrimination
since classes of DE models can be placed in this plane. Comparing data models we find that  
some degeneracies among cosmological models remain even for  
the most ambitious project.  
Complementary information is contained in a  
representation of $w(z)$ with its error shape. This allows us to understand  
the compatibility of the model with the different observables,
and in particular, to represent the redshift dependence of the error.
However,
the results remain in general parameterization dependent and the interpretation 
is challenging.
This redshift plane may be useful for consistency checks and 
data model compatibility and comparison.\\

A solution to improve the sensitivity of the SN analysis is  to combine 
SN data with other probe information. This is certainly  powerful as emphasized by the 
DETF but this should be manipulated with  some caution as systematic errors will 
dominate the future analyses  and will introduce even stronger bias in a combination. 
The best test  will be to check the compatibility between probes when dominated by  
systematic errors, in a coherent way (same theoretical assumptions,  same framework, 
same treatment of systematic errors). Combination of  probes two by two will then 
help to control systematic effects. This  will be also a good cross check of 
the internal hypothesis and a control of results.\\

\no {\bf Acknowledgments}\\ 

Special thanks to A. Tilquin, P. Taxil, C. Marinoni, C. Tao,
D. Fouchez and A. Bonissent for helpful comments and suggestions.
We also thank the members of the ``Laboratoire d'Astrophysique de Marseille''
for interesting discussions, in particular A. Mazure, J.-P. Kneib and R. Malina.

\label{lastpage}

\end{document}

%% file: macros.tex
\def\eg{{e.g.,~}}
\def\ie{{i.e.,~}}
\def\lsim{\raise0.3ex\hbox{$<$}\kern-0.75em{\lower0.65ex\hbox{$\sim$}}} 
\def\gsim{\raise0.3ex\hbox{$>$}\kern-0.75em{\lower0.65ex\hbox{$\sim$}}} 
\def\lesssim{\mathrel{\hbox{\rlap{\hbox{\lower4pt\hbox{$\sim$}}}\hbox{$<$}}}}
\def\gtrsim{\mathrel{\hbox{\rlap{\hbox{\lower4pt\hbox{$\sim$}}}\hbox{$>$}}}}
\def\sc1{\raise2.1ex\hbox{\tiny $r\!\!=\!\!4$}\kern-0.95em{\hbox{$=$}}}

\newcommand{\unit}[1]{\ifmmode \:\mbox{\rm #1}\else \mbox{#1}\fi}

\def\ltsima{$\; \buildrel < \over \sim \;$}
\def\simlt{\lower.5ex\hbox{\ltsima}}
\def\gtsima{$\; \buildrel > \over \sim \;$}
\def\simgt{\lower.5ex\hbox{\gtsima}}          

\def\sc{{\rm Science\ }}